\documentclass[11pt,letterpaper]{article}
\pdfoutput=1

\usepackage{latexsym}
\usepackage{epsfig}
\usepackage{verbatim}
\usepackage{multirow}
\usepackage{fancybox}
\usepackage{shadow}
\usepackage{cite}
\usepackage{amscd, amsmath, amssymb}
\usepackage{bm}
\usepackage{txfonts, yfonts, pmboxdraw}
\usepackage{graphicx}

\newtheorem{theorem}{\sc Theorem}[section]
\newtheorem{lemma}[theorem]{\sc Lemma}
\newtheorem{corollary}[theorem]{\sc Corollary}
\newtheorem{fact}[theorem]{\sc Fact}

\newcommand{\eps}{\varepsilon}

\newcommand{\proofend}{{\medskip\medskip}}
\newcommand{\proof}{{\noindent\em Proof. }}

\author{
  {\sc Bernard Chazelle}
\thanks{Department of Computer Science,
       Princeton University, 
{\tt chazelle}@{\tt cs.princeton.edu }}
}

\title{Toward a Theory of Markov Influence Systems 
and their Renormalization\thanks{A preliminary version of this work appeared in
the {\em Proceedings of the
9th Innovations in Theoretical Computer Science (ITCS), 2018.}
The Research was sponsored by the Army
Research Office and the Defense
Advanced Research Projects Agency and was accomplished under Grant Number
W911NF-17-1-0078.
The views and conclusions contained in this document are those of the
authors and should not be
interpreted as representing the official policies, either expressed or
implied, of the Army Research Office,
the Defense Advanced Research Projects Agency, or the U.S. Government. The
U.S. Government is
authorized to reproduce and distribute reprints for Government purposes
notwithstanding any copyright
notation herein.
}}

\date{}

\begin{document} \maketitle

\vspace{.5cm}

\begin{abstract}
We introduce the concept of
a {\em Markov influence system} ({\em MIS})
and analyze its dynamics. An {\em MIS} 
models a random walk in a graph whose edges and
transition probabilities change endogenously as a function of the
current distribution.  This article consists of two independent parts:
in the first one, 
we generalize the standard classification of Markov chain states to the
time-varying case
by showing how to ``parse" graph sequences;
in the second part, we use this framework to
carry out the bifurcation analysis of a few important {\em MIS} families.
We show that, in general, these systems can be chaotic but that 
irreducible {\em MIS} are almost always
asymptotically periodic. We give an example of ``hyper-torpid" mixing,
where a stationary distribution is reached in super-exponential time,
a timescale beyond the reach of any Markov chain.
\end{abstract}

\vspace{.5cm}

\noindent
{\small
{\em Keywords:}
Random walks on time-varying graphs; Markov influence systems; Graph sequence parsing;
Renormalization; Hyper-torpid mixing; Chaos
}

\section{Introduction}

Nonlinear Markov chains are popular probabilistic models
in the natural and social sciences. They are commonly used in
interacting particle systems, epidemic models, replicator dynamics,
mean-field games, etc.~\cite{caoSM, coppdiac1, coppdiac2, frank2013, IacobFRW, KolokBook, seneta06}.
They differ from the linear kind by
allowing transition probabilities to vary as a function of the current
state distribution.
For example, a traffic network might 
update its topology and edge transition rates adaptively
to alleviate congestion.  
The systems are Markovian in that the future
depends only on the present: in this work,
the present will refer to the current state distribution 
rather than the single state presently visited.
The traditional formulation of these
models comes from physics and relies on the classic tools of the trade:
stochastic differential calculus, McKean interpretations, Feynman-Kac models, 
Fokker-Planck PDEs, etc.~\cite{castellanoFL2009, chazWellPosed, frank2013, KolokBook}.
These techniques assume symmetries
that are typically absent from the ``mesoscopic" scales of {\em natural algorithms};
they also often operate at the thermodynamic limit, which rules out genuine
agent-based modeling. Our goal is to initiate a theory of
discrete-time Markov chains whose topologies vary
as a function of the current probability distribution. Thus the entire
theory of finite Markov chains should be recoverable as a special case.
Our contribution comes in two independent parts:
the first one is a generalization of the classification of Markov chains to the
time-varying case; the second part is the bifurcation analysis of
Markov influence systems. The work highlights the signature trait of
time-varying random walks, which is the possibility
of super-exponential (``hyper-torpid") mixing.

\paragraph{Renormalization.}

The term refers to a wide-reaching approach to complex systems that
originated in quantum field theory and later expanded into
statistical mechanics and dynamics.
Whether in its exact or coarse-grained form, the basic idea
is to break down a complex system into
a hierarchy of simpler parts. 
When we define a dynamics on the original system (think of 
interacting particles moving about) then the hierarchy itself creates 
its own dynamics between the layers. This new ``renormalized" dynamics can be entirely
different from the original one. Crucially, it can be both easier to analyze
and more readily expressive of global properties. For example, 
second-order phase transitions in the Ising model might correspond to
fixed points of the renormalized dynamics.

%%%   \footnote{The idea is
%%%   very powerful:  Ken Wilson won the 1982 Nobel
%%%   prize in physics and Artur Avila the 2014 Fields medal for their 
%%%   (very different) breakthroughs in the use of renormalization:
%%%   finding new critical exponents; proving the weak mixing of interval exchange transformations, etc.}

\begin{figure}[htb]
\begin{center}
\hspace{0cm}
\includegraphics[width=6cm]{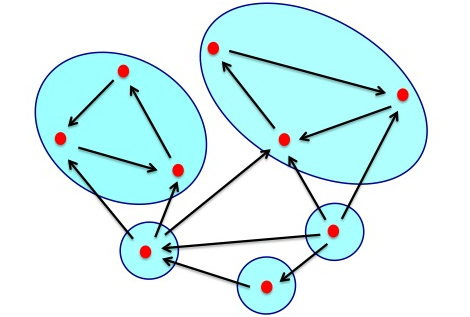}
\end{center}
\vspace{-0.4cm}
\caption{\small  The condensation of a graph. \label{label-misFig1}}
\end{figure}

What is the relation to Markov chains?  
The standard classification of
the states of a Markov chain is an example of exact renormalization.
Recall that the main idea behind the classification is to express the chain as an acyclic directed graph,
its {\em condensation}, whose vertices correspond to its strongly connected components.
This creates a two-level hierarchy~(fig.\ref{label-misFig1}): a tree with a root (the condensation) and its children
(the strongly connected components).
In a random walk, the 
probability mass will flow entirely into the $k\,\,  (=2)$ sinks of the condensation.
The stationary distribution lies on an attracting manifold of
dimension $k-1\, \, (=1)$.
The hierarchy has only two levels.
The situation is different with time-varying Markov chains, 
which can have deep hierarchies.

\vspace{0.1cm}

Consider an infinite sequence $(g_k)_{k>0}$ of digraphs over a fixed 
set of labelled vertices. A {\em temporal} random walk is defined by
picking a starting vertex, moving to a random neighbor in~$g_1$,
then a random neighbor in $g_2$, and so on, 
forever~\cite{condonhernek94, CondonL, holme-saram, OthonM, starniniBBP2012}.
Note that a temporal walk might not match a path in any of the graphs.
How would one classify the states of this ``dynamic" Markov chain?
Repeating the condensation decomposition at each step makes little sense,
as it carries no information about the temporal walks.
Instead, we want to monitor when and where 
temporal walks are {\em extended}. The {\em cumulant} graph
compiles all such extensions and,
when the process stalls at time $t'$, reboots it by restarting from scratch 
at $t=t'$.  
We define a {\em grammar} with which we
parse the sequence $(g_k)_{k>0}$ accordingly.
The method, explained in detail in the next section,
is very general and likely to be useful elsewhere.

\paragraph{Markov influence systems.}

All finite Markov chains oscillate periodically or mix to a stationary distribution. 
One key fact about their dynamics is that the timescales never exceed a single exponential
in the number of states.  Allowing the transition probabilities to 
fluctuate over time at random does not change that 
basic fact~\cite{avinKL, DenysyukR11, DenysyukR14}.
Markov influence systems are quite different in that regard.
Postponing formal definitions, let us think of an {\em MIS} for now as
a dynamical system defined by iterating the map 
$f\! : \mathbf{x}^\top\mapsto \mathbf{x}^\top S(\mathbf{x})$,
where $\mathbf{x}$ is a probability distribution represented as a column vector in
$\mathbb{R}^n$ and $S(\mathbf{x})$
is a stochastic matrix that is piecewise-constant as a function of $\mathbf{x}$.
We assume that the discontinuities are flats (ie, affine subspaces).  
The assumption is not nearly as restrictive as it appears,
as we explain with a simple example.

Consider a random variable $\xi$ over the distribution $\mathbf{x}$
and fix two $n$-by-$n$ stochastic matrices $A$ and~$B$.
Define $S(\mathbf{x})= A$ (resp. $B$) if $\text{\rm var}_\mathbf{x}\, \xi>1$ (resp. else);
in other words, the random walk picks one of two stochastic matrices at each step
depending on the variance of $\xi$ with respect to the current state distribution $\mathbf{x}$.
Note that, in violation of our assumption,
the discontinuity is quadratic in $\mathbf{x}$.
This is not an issue because we can always linearize the variance:
we begin with the identity
$\text{\rm var}_\mathbf{x}\, \xi= \frac{1}{2} \sum_{i,j}\, (\xi_i-\xi_j)^2 x_ix_j$
and the fact that $\mathbf{y}\! := (x_ix_j)_{i,j}$ is a probability distribution.
We form the Kronecker square
$T(\mathbf{y})= S(\mathbf{x})\otimes S(\mathbf{x})$ and lift
the system into the $(n^2-1)$-dimensional unit simplex to get a brand-new {\em MIS}
defined by the map $\mathbf{y}^\top \mapsto \mathbf{y}^\top T(\mathbf{y})$. We now have 
linear discontinuities.
This same type of tensor lift can be used to linearize 
any algebraic constraints; this requires making the polynomials
over $x_i$ homogeneous, which we can do by using the identity $\sum_i x_i=1$.
Using ideas from~\cite{chaz-dis-15}, one can go even further than that
and base the stepwise edge selection on the outcome
of any first-order logical formula we may fancy 
(with the $x_i$'s acting as free variables). The key fact behind this result
is that the first-order theory of the reals is decidable by quantifier elimination.
This allows us to pick the next stochastic matrix at each time step
on the basis of the truth value of a Boolean logic formula
with arbitrarily many quantifiers (see~\cite{chaz-dis-15} for details).
This discussion is meant to highlight the fact that
assuming linear discontinuities is not truly restrictive. 

We prove in this article that an irreducible Markov influence system is almost always
asymptotically periodic. (An {\em MIS} is irreducible
if $S(\mathbf{x})$ forms an irreducible chain for each $\mathbf{x}$.)
We extend this result to larger families of Markov influence systems.
We also give an example of ``hyper-torpid" mixing: an {\em MIS} that
converges to a stationary distribution in time equal to a tower-of-twos
in the size of the chain. This bound also applies to the period of 
certain periodic {\em MIS}.
The emergence of timescales far beyond the reach of
standard Markov chains is a distinctive feature of Markov influence systems.

Unlike in a standard random walk, the noncontractive eigenspace
of an {\em MIS} may vary over time. 
It is this spectral incoherence that renormalization attempts to ``tame."
To see why this has a strong graph-theoretic flavor, observe that
at each time step the support of the stationary distribution can be read off 
the topology of the current graph:
for example, the number of sinks in the condensation is equal to
the dimension of the principal eigenspace plus one. Renormalization can thus
be seen as an attempt to restore coherence to an ever-changing 
spectral landscape via a dynamic hierarchy of graphs, subgraphs, and homomorphs.

The bifurcation analysis at the heart of the analysis
depends on a notion of 
``general position" aimed at bounding the growth rate
of the induced symbolic 
dynamics~\cite{bruinD09, katokHassel, sternberg10}.
The root of the problem is a clash between order and
randomness similar to the conflict 
between entropy and energy encountered in the Ising model.
The tension between these two ``forces" is  
mediated by the critical values of a perturbation parameter,
which are shown to form 
a Cantor set of Hausdorff dimension strictly less than~1.
Final remark:
There is a growing body of literature on dynamic 
graphs~\cite{AKMUV12, barratBook08, 
castellanoFL2009, fagnaniBook,
holme-saram, KKK2000, OthonM, MichailS,
ProskTempoI, ProskTempoII}
and their random walks~\cite{avinKL, condonhernek94, CondonL, coppdiac1, 
coppdiac2, DenysyukR11, DenysyukR14, IacobFRW, NguyenEtAl18,
PerraBMGPV, ramiroLSR, starniniBBP2012}.
What distinguishes this work from its predecessors is that
the changes in topology are induced endogenously 
by the system itself (via a feedback loop).

\section{How to Parse a Graph Sequence}

Throughout this work, a {\em digraph} refers to a directed graph with vertices
in $[n]\!: = \{1,\ldots, n\}$ and a self-loop at each vertex.
Graphs and digraphs (words we use interchangeably) have no  
multiple edges.
We denote digraphs by lower-case letters ($g,h$, etc) and use boldface
symbols for sequences.
A {\em digraph sequence} $\mathbf{g}= (g_k)_{k> 0}$
is an ordered, finite or infinite, list of digraphs over the vertex set $[n]$.
The digraph $g_i\times g_j$ consists of all the edges $(x,y)$ such that
there exist an edge $(x,z)$ in $g_i$ and another one $(z,y)$ in $g_j$
for at least one vertex $z$.
The operation $\times$ is associative but not commutative;
it corresponds roughly to matrix multiplication.
We define the {\em cumulant} $\prod_{\leq k} \mathbf{g}=g_1\times \cdots \times g_k$ 
and write $\prod \mathbf{g}= g_1 \times g_2\times \cdots$
for finite $\mathbf{g}$. 
The cumulant indicates all the pairs of vertices
that can be joined by a temporal walk of a given length.
The mixing time of a random walk on a (fixed) 
graph depends on the speed
at which information propagates and, in particular, 
how quickly the cumulant becomes transitive. In the time-varying case,
mixing is a more complicated proposition, but the emergence  
of transitive cumulants is still what guides the parsing process.

An edge $(x,y)$ of a digraph $g$ is {\em leading} if
there is $u$ such that $(u,x)$ is an edge of $g$ but $(u,y)$ is not.
The non-leading edges form a subgraph of $g$, 
which is denoted by $\text{\em{tf}}\,(g)$ and called the transitive front of $g$.
For example,
$\text{\em{tf}}\,(x\rightarrow y\rightarrow z)$ is the graph
over $x,y,z$ with the single edge $x\rightarrow y$ (and the three self-loops);
on the other hand, 
the transitive front of a directed cycle over three or more vertices
has no edges besides the self-loops.
We denote by $\text{\em{cl}} (g)$ 
the transitive closure of $g$: it is the graph that includes an edge
$(x,y)$ for any two vertices $x,y$ with a path from $x$ to $y$.
Note that $\text{\em{tf}}\, (g)\preceq g \preceq \text{\em{cl}} (g)$.

\begin{itemize}
\item
An equivalent definition of the transitive front is that the edges of
$\text{\em{tf}}\,(g)$ are precisely the pairs $(i,j)$ such that
$C_i\subseteq C_j$, where $C_k$ denotes the set of vertices $l$ 
such that $(l,k)$ is an edge of $g$.
Because each vertex has a self-loop, the inclusion $C_i\subseteq C_j$
implies that $(i,j)$ is an edge of $g$.
If $g$ is transitive, then $\text{\em{tf}}\,(g)= g$. 
The set-inclusion definition of the transitive front
shows that it is indeed transitive:
ie, if $(x,y)$ and $(y,z)$ are edges, then so is $(x,z)$.
Given two graphs $g, h$ over the same vertex set,
we write $g \preceq h$ if all the edges of $g$ are in $h$
(with strict inclusion denoted by the symbol $\prec$).
Because of the self-loops, $g, h \preceq g\times h$. 

\item
A third characterization of
$\text{\em{tf}}\,(g)$ is as the unique densest graph $h$ over $[n]$ such that $g\times h=g$:
we call this the {\em maximally-dense property} of the transitive front,
and it is the motivation behind our use of the concept.
Indeed,  the failure of subsequent graphs to grow
the cumulant implies a structural constraint on them.
This is the sort of structure that parsing attempts to tease out.
\end{itemize}

\paragraph{The parser.}

The parse tree of a (finite or infinite) graph sequence $\mathbf{g}= ( g_k)_{k> 0}$
is a rooted tree whose leaves are associated with $g_1,g_2,\ldots$ from left to right.
The purpose of the parse tree is to monitor
the formation of new temporal walks as time progresses.
This is based on the observation that, because of the self-loops, the cumulant
$\prod_{\leq k} \mathbf{g}$ is monotonically nondecreasing with $k$
(with all references to graph ordering being relative to $\preceq$).
If the increase were strict at each step then the parse tree would look
like a fishbone. 
But, of course, the increase cannot go on forever.
To continue to extract structure even when the cumulant is ``stuck" is 
what parsing is all about.
The underlying grammar consists of three productions:
(1a) and~(1b) renormalize the graph sequence along the time axis,
while (2) creates the  hierarchical clustering of the graphs in the sequence~$\mathbf{g}$.

\begin{enumerate}
\item
{\sc Temporal renormalization} \hspace{.3cm}
We express the sequence $\mathbf{g}$ in terms
of minimal subsequences with cumulants equal to $\prod \mathbf{g}$.
There is a unique decomposition
$$\mathbf{g} =
\mathbf{g}_1, g_{m_1}, \ldots,
\mathbf{g}_k, g_{m_k},
\mathbf{g}_{k+1}
$$
such that
\begin{itemize}
\item[(i)]
$\mathbf{g}_1= g_{1},\ldots, g_{m_1-1}$;
$\mathbf{g}_i= g_{m_{i-1}+1},\ldots, g_{m_i-1}$
($1< i\leq k$); and $\mathbf{g}_{k+1}= g_{m_k+1},\ldots\,$.
\item[(ii)]
$\bigl( \prod \mathbf{g}_i\bigr)\times g_{m_i} = \prod \mathbf{g}$, for any $i\leq k$;
and $\prod \mathbf{g}_i \prec \prod \mathbf{g}$, for any $i\leq k+1$.
\end{itemize}
The two productions below create the {\em temporal parse tree}.

\begin{itemize}
\item
{\em Transitivization.} 
Assume that $\prod \mathbf{g}$ is not transitive.
We define $h= \text{\em{tf}}\,(\prod \mathbf{g})$
and note that $h\prec \prod \mathbf{g}$.
It follows from the
maximally-dense property of the transitive front that $k=1$.
Indeed, $k>1$ implies that 
$\prod \mathbf{g} =
\bigl( \prod \mathbf{g}_2\bigr)\times  g_{m_2}\preceq 
 \text{\em{tf}}\, \bigl\{ \bigl( \prod \mathbf{g}_1\bigr)\times g_{m_1} \bigr\}= 
  \text{\em{tf}}\,(\prod \mathbf{g})$,
which contradicts the non-transitivity of $\prod \mathbf{g}$.
We have the production
\begin{equation}\label{prod1a}
\mathbf{g}
    \,\,  \longrightarrow \,\,
    \bigl(\, \mathbf{g}_1 \bigr)\, g_{m_1}\, 
    \bigl(\, (\mathbf{g}_2 ) \bigtriangleup  h \, \bigr) ,
\tag{1a}
\end{equation}
In the parse tree, the node for $\mathbf{g}$ has three children: the first
one serves as the root of the temporal parse subtree for $\mathbf{g}_1$;
the second one is the leaf associated with the graph $g_{m_1}$; the
third one is a {\em special} node annotated with the label $h$, which serves as the parent $v$
of the node $w$ rooting the parse subtree for $\mathbf{g}_2$. 
The node $w$ is labeled (implicitly) by the transitive
graph $\text{\em{cl}} (\prod \mathbf{g}_2)$,
which serves as a coarse-grained approximation of $\prod \mathbf{g}_2$.
The purpose of annotating a special node with the label $h$ 
is to provide an intermediate approximation of $\prod \mathbf{g}_2$
that is strictly finer than the ``obvious" $\text{\em{cl}} (\prod \mathbf{g})$.
These coarse-grained approximations are called {\em sketches}.

\item
{\em Cumulant completion.} 
Assume that $\prod \mathbf{g}$ is transitive. We have the production
\begin{equation}\label{prod1b}
\mathbf{g}
    \,\,  \longrightarrow \,\,
    \bigl(\, \mathbf{g}_1 \bigr)\, g_{m_1}\, 
    \bigl(\, \mathbf{g}_2   \bigr)\, g_{m_2}\, 
    \cdots 
    \bigl(\, \mathbf{g}_k  \bigr)\, g_{m_k}\, 
    \bigl(\, \mathbf{g}_{k+1} \bigr).
\tag{1b}
\end{equation}
Note that the index $k$ may be infinite and
any of the subsequences $\mathbf{g}_i$ might be empty
(for example, $\mathbf{g}_{k+1}$ if $k= \infty$).
\end{itemize}

\item
{\sc Topological renormalization}  \hspace{.3cm}
Network renormalization exploits the fact that 
the information flowing across the system might get stuck in portions of the graph
for some period of time: when this happens, 
we cluster the graphs using topological renormalization.
As we just discussed, each
node $v$ of the temporal parse tree that is not annotated by~(\ref{prod1a})
is labeled by the sketch $\text{\em{cl}} (\prod \mathbf{g})$, 
where $\mathbf{g}$ is the graph sequence formed by the leaves of the subtree rooted
at $v$. In this way, every path from the root of the temporal parse tree
comes with a nested sequence of sketches  
$h_1\succeq \cdots \succeq h_l$. Pick two consecutive ones, $h_i, h_{i+1}$:
these are two transitive graphs whose strongly connected components, therefore, are cliques.
Let $V_1,\ldots, V_a$ and $W_1,\ldots, W_b$ be the vertex sets of the
cliques corresponding to $h_i$ and $h_{i+1}$, respectively. Since $h_{i+1}$ is a subgraph
of $h_i$, it follows that each $V_i$ is a disjoint union of the form
$W_{i_1}\cup \cdots \cup W_{i_{s_i}}$. 

\begin{itemize}
\item
{\em Decoupling.} 
We decorate the temporal parse tree with an additional tree connecting
the sketches present along each one of its paths.
These {\em topological parse trees} are formed by all the productions of the type:
\begin{equation}\label{prod2}
V_i 
    \,\,  \longrightarrow \,\,
    W_{i_1} \cdots W_{i_{s_i}}.
\end{equation}
A sketch at a node $v$ of the temporal tree 
can be viewed as an acyclic digraph over cliques: its
purpose is to place limits on
the movement of the probability mass in any temporal random walk
corresponding to the leaves of the subtree rooted at $v$.
In particular, it indicates how decoupling might arise in the system.
\end{itemize}

\end{enumerate}

The maximum depth of the temporal parse tree is $O(n^2)$ 
because each child's cumulant loses at least one edge 
from its parent's (or grandparent's) cumulant.
To see why the quadratic bound is tight,
consider a bipartite graph $V=L\cup R$, where $|L|= |R|$ and
each pair from $L\times R$ is added one at a time as a bipartite graph with a single nonloop edge;
the leftmost path of the parse tree is of quadratic length.

\paragraph{Left-to-right parsing.}

The temporal tree can be built on-line by scanning the graph sequence $\mathbf{g}$ 
with no need to back up.
Let $\mathbf{g'}$ denote the sequence formed by appending
the graph $g$ to the end of the finite graph sequence $\mathbf{g}$.
If $\mathbf{g}$ is empty, then the tree $\mathcal{T}(\mathbf{g'})$ consists
of a root with one child labeled $g$.
If $\mathbf{g}$ is not empty and
$\prod \mathbf{g}\prec \prod \mathbf{g'}$,
the root of $\mathcal{T}(\mathbf{g'})$ has one child
formed by the root of $\mathcal{T}(\mathbf{g})$ as well as a
child labeled $g$. Assume now that $\mathbf{g}$ is not empty
and that $\prod \mathbf{g}= \prod \mathbf{g'}$.
Let $v$ be the lowest internal node on the rightmost path
of $\mathcal{T}(\mathbf{g})$ such that $c_v\times g= c_v$, where
$c_u$ denotes the product of the graphs associated with the leaves
of the subtree rooted at node $u$ of $\mathcal{T}(\mathbf{g})$.
Let $w$ be the rightmost child of $v$; note
that $v$ and $w$ always exist. We explain how to form 
$\mathcal{T}(\mathbf{g'})$ by editing $\mathcal{T}(\mathbf{g})$.

\begin{enumerate}
\item
$c_v$ is transitive and $w$ is a leaf:
We add a leaf labeled $g$ as the new rightmost child $z$ of $v$ 
if $g=c_v$; otherwise, we do the same but, instead
of attaching $g$ directly to $v$, we make it the unique child of~$z$.

\item
$c_v$ is transitive and $w$ is not a leaf:
We add a leaf labeled $g$ as the new rightmost child of $v$
if $c_w\times g = c_v$; otherwise, 
we add a new rightmost child $z$ to $v$ and we give $z$
two children: its left child is $w$ and its subtree below;
its right child is a leaf labeled $g$.

\item
$c_v$ is not transitive and $w$ is a leaf:
We attach a new rightmost child $z$ to $v$,
and we make it a special node annotated with the label $\text{\em{tf}}\,(c_v)$.
We give $z$ a single child $z'$, which is itself the parent of a new leaf
labeled $g$.

\item
$c_v$ is not transitive and $w$ is not a leaf:
We note that $c_w\times g  \preceq \text{\em{tf}}\,(c_v) \prec c_v$.
We attach a new rightmost child $z$ to $v$,
and we make it a special node annotated with the label $\text{\em{tf}}\,(c_v)$.
We give $z$ a single child $z'$, to which we attach
two children: one of them is $w$ (and its subtree)
and the other one is the leaf labeled $g$.
\end{enumerate}

\paragraph{Undirected graphs.}

A graph is called undirected if any edge $(x,y)$ with $x\neq y$
comes with its companion $(y,x)$. Consider a sequence of undirected graphs over $[n]$.
We begin with the observation that the cumulant of a subsequence 
might itself be directed; for example, the product
$g_1\times g_2= (x\leftrightarrow y \ \ \ \ z) \times (x \ \ \ \ y\leftrightarrow z)$
has a directed edge from $x$ to $z$ but not from $z$ to $x$.
We can use undirectedness to strengthen the definition of the transitive front.
Recall that $\text{\em{tf}}\,(g)$ is the unique densest graph $h$ such that $g\times h=g$.
Its purpose is the following:
if $g$ is the current cumulant, the transitive front of $g$
is intended to include any edge that might appear in subsequent graphs in the sequence
without extending any path in $g$.
Since, in the present case, the only edges
considered for extension will be undirected, we might as well require that $h$ itself
(unlike $g$) should be undirected.
In this way, we redefine the transitive front, now denoted by $\text{\em{utf}}\,(g)$,
as the unique densest {\em undirected} graph $h$ such that $g\times h=g$.
Its edge set includes all the pairs $(i,j)$ such that $C_i= C_j$.
Because of self-loops, the condition implies that $(i,j)$ is
an undirected edge of $g$. This forms an equivalence relation among the vertices,
so that $\text{\em{utf}}\,(g)$ actually consists of disconnected, undirected cliques. 
To see the difference with the directed case, we take our
previous example and note that 
$\text{\em{tf}}\,(g_1\times g_2)$ has the
edges $(x,y), (x,z), (y,z), (z,y)$ in addition to the self-loops, whereas
$\text{\em{utf}}\,(g_1\times g_2)$ has the single undirected edge $(y,z)$ plus
self-loops.

The depth of the parse tree can still be as high
as quadratic in $n$. To see why, consider the following recursive
construction. Given a clique $C_k$ over $k$ vertices $x_1,\ldots, x_k$
at time $t$, attach to it, at time $t+1$, the undirected edge $(x_1,y)$.
The cumulant gains the undirected edge $(x_1,y)$ and the 
directed edges $(x_i,y)$ for $i=2,\ldots, k$.
At time $t+2,\ldots, t+k$, visit each one of the $k-1$ undirected edges
$(x_1,x_i)$ for $i>1$, using single-edge undirected graphs.
Each such step will see the addition of a new directed edge $(y,x_i)$ to the cumulant, until it becomes
the undirected clique $C_{k+1}$. The quadratic lower bound on the tree depth follows immediately.

\paragraph{Backward parsing.}

The sequence of graphs leads to products where each new graph is
multiplied to the right, as would happen in a 
time-varying Markov chain. Algebraically, the matrices
are multiplied from left to right. In diffusive systems
(eg, multiagent agreement systems, Hegselmann-Krause models, 
Deffuant systems), however, matrices are multiplied from right to left.
Although the dynamics can be 
quite different, the same parsing algorithm can be used.
Given a sequence $\mathbf{g}= (g_k)_{k> 0}$,
its {\em backward parsing} is formed by applying the parser to the sequence
$\overleftarrow{\,\mathbf{g}} = (h_k)_{k> 0}$, where $h_k$ is derived from $g_k$ by
reversing the direction of every edge, ie,
$(x,y)$ becomes $(y,x)$.
Once the parse tree for $\overleftarrow{\,\mathbf{g}}$
has been built, we simply restore each edge
to its proper direction to produce the {\em backward parse tree} of $\mathbf{g}$.

\begin{comment}
<<<<<<<<<<<<<<<<<<<<<<<<<<<<<<<<<<<<<<<<<<<<<<<<<<<<
\paragraph{Discussion.}

For the chronological ordering of the graphs in the sequence
to match the left-to-right order of the leaves in the parse tree, it helps to
think of the tree as a sort of Calder mobile in 3D where the 
network renormalization nodes~(\ref{prod2a}, \ref{prod2b})
are presented along an axis normal to the page. 
The grammar parses any sentence in the language of all graphs
on $n$ vertices. It does so by scanning through the sequence once
with no need to back up. Indeed, the parse tree for any prefix
of $\mathbf{g}$ is a subtree of the parse tree for $\mathbf{g}$.
Renormalization in ``network space" and ``time space"
builds a coherent picture of the graph evolution at different scales.
Intuitively, the graphs are clustered hierarchically
in such a way that the top layers conform to 
the accumulation of graphs far in the past while the bottom
of the hierarchy reflects more recent trends.   Thus levels in 
network space match timescales, so that rapid changes in
the graph sequence over short time windows translate into quick updates
of clusters down the hierarchy.
>>>>>>>>>>>>>>>>>>>>>>>>>>>>>>>>>>>>>>>>>>>>>>>>>>>>>
\end{comment}

\section{The Markov Influence Model}\label{MIS}

\begin{figure}[htb]
\begin{center}
\hspace{-0.5cm}
\includegraphics[width=7cm]{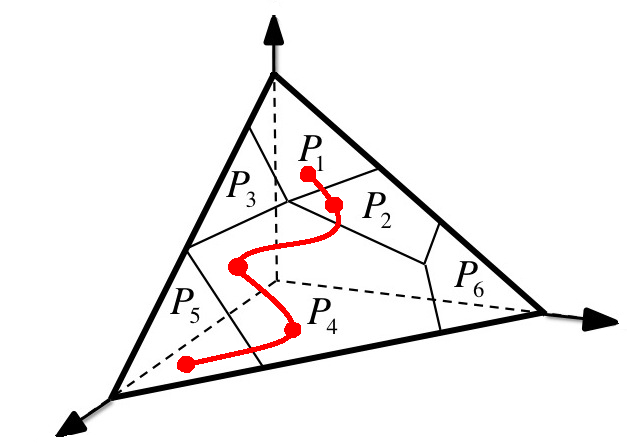}
\end{center}
\vspace{-0.5cm}
\caption{\small  
A Markov influence system: each region $P_k$ is
associated with a single stochastic matrix. 
The first five steps of an orbit are shown to visit
regions $P_1,P_2,P_4,P_4,P_5$ in this order. 
\label{label-misFig2}}
\end{figure}

Let $\mathbb{S}^{n-1}$ (or $\mathbb{S}$ when the dimension is understood) 
be the standard simplex $\bigl\{\, \mathbf{x}\in \mathbb{R}^n\,|\,\,
\mathbf{x}\geq \mathbf{0} \, , \,  \|\mathbf{x}\|_1=1 \,\bigr\}$
and let $\mathcal{S}$ denote set of all $n$-by-$n$ rational stochastic matrices. 
A {\em Markov influence system} ({\em MIS\,}) 
is a discrete-time dynamical system with phase space $\mathbb{S}$, which is
defined by the map $f\! : \mathbf{x}^\top \mapsto f(\mathbf{x})\! := \mathbf{x}^\top S\!(\mathbf{x})$,
where $\mathbf{x}\in \mathbb{S}$
and $S$ is a function $\mathbb{S}\mapsto \mathcal{S}$
that is constant over the pieces of a finite polyhedral partition
$\mathcal{P}\! = \! \{P_k\}$ of $\mathbb{S}$;
we define $f$ as
the identity on the discontinuities of the partition~(fig.\ref{label-misFig2}).

\smallskip
{\em{(i)}}\ \ 
We define the digraph $g(\mathbf{x})$
(and its corresponding Markov chain) formed by the positive entries of $S\!(\mathbf{x})$.
To avoid inessential technicalities, we assume that the diagonal of each
$S\!(\mathbf{x})$ is strictly positive (ie, $g(\mathbf{x})$ has self-loops).
In this way, any orbit of an {\em MIS} corresponds 
to a lazy, time-varying random walk with transitions defined endogenously.\footnote{As 
discussed in the introduction, to access
the full power of first-order logic in the stepwise choice of digraphs requires nonlinear
partitions, but these can be linearized by a suitable tensor construction~\cite{chaz-dis-15}.}
We recall some basic terminology.
The {\em orbit} of $\mathbf{x}\in \mathbb{S}$ is the infinite sequence 
$(f^t(\mathbf{x}))_{t\geq 0}$ and its {\em itinerary} is the corresponding
sequence of cells $P_k$ visited in the process.  The orbit is {\em periodic}
if $f^t(\mathbf{x})= f^{s}(\mathbf{x})$ for any $s=t$ modulo a fixed integer.
It is asymptotically periodic if it gets arbitrarily close to a periodic orbit over time.

\vspace{0.0cm}

\smallskip
{\em{(ii)}}\ \ 
The discontinuities in $\mathcal{P}$ are formed by hyperplanes 
in $\mathbb{R}^n$ of the form $\mathbf{a}_i^\top \mathbf{x} =1+\delta$,
where $\delta$ is confined to an interval $\Omega:= [-\omega, \omega]$.
Assuming general position, we can pick a small positive $\omega<1/2$ 	
so that $\mathcal{P}$ remains (topologically) the same 
over $\Omega$.\footnote{Let $H_\delta$ be the set
consisting of the hyperplane $\sum_i x_i=1$ together with those used
to define $\mathcal{P}$.  We assume that $H_0$ is in general position; hence
so is $H_\delta$ for any $\delta\in \Omega$, where $0<\omega<1/2$ is smaller than
a value that depends only on the set of vectors $\{\mathbf{a}_i\}$.
Note that this problem arises only because of the constraint $\sum_i x_i=1$,
since otherwise the hyperplane arrangement is central around $(0,\ldots,0, -1)$.}
Thus, the {\em MIS} remains well-defined for all $\delta\in \Omega$.
As we show in Section~\ref{chaos}, 
the parameter $\delta$ is necessary to keep chaos at bay.

\smallskip
{\em{(iii)}}\ \ 
The {\em coefficient of ergodicity} $\tau(M)$ of a matrix $M$
is defined as half the maximum $\ell_1$-distance between any two of its 
rows~\cite{seneta06}. It is submultiplicative for stochastic matrices,
a direct consequence of the identity  
%%%
%%%    SENETA     Lemma 4.3 on page 139.
%%%
$\tau(M)= \max \,\Bigl\{\, \|\mathbf{x}^\top M\|_1 :  \mathbf{x}^\top \mathbf{1}=0
\hspace{.2cm}\text{and}\hspace{.2cm} \|\mathbf{x}\|_1=1\,\Bigr\}$.

\smallskip
{\em{(iv)}}\ \ 
Given $\Delta\subseteq \Omega$, let $L_\Delta^t$
denote the set of $t$-long prefixes of any itinerary
for any starting position $\mathbf{x}\in \mathbb{S}$
and any $\delta\in \Delta$.
We define the {\em ergodic renormalizer} $\eta_\Delta$ as the
smallest integer such that, for any $t\geq \eta_\Delta$
and any matrix sequence  $S_{\!1},\ldots, S_{\! t}$ associated with
an element of $L_\Delta^t$,
the product $S_{\! 1}\cdots S_{\! t}$ is primitive (ie, some high
enough power is a positive matrix)
and its coefficient of ergodicity is less than~$1/2$.
We assume in this section 
that $\eta \! := \eta_\Omega <\infty$ and discuss in Section~\ref{apps}
how to relax this assumption via renormalization.
Let $D$ be the union of the hyperplanes from 
$\mathcal{P}$ in $\mathbb{R}^n$ (where $\delta$ is understood).
We define $Z_t= \bigcup_{0\leq k\leq t}f^{-k}(D)$
and $Z= \bigcup_{t\geq 0} Z_t$.
Note that $Z\subseteq \mathbb{S}$ 
since the latter is the domain of $f$.
Remarkably, for almost all $\delta\in \Omega$,
$Z_t$ becomes strictly equal to $Z$ in a finite number of steps.\footnote{Recall
that both $Z$ and $Z_t$ depend on $\delta\in \Omega$.
All the constants used in this work
may depend on the system's parameters such as $n$, $\mathcal{P}$
(but not on $\delta$).
Dependency on other parameters is indicated by a subscript.}

\begin{lemma}\label{MIS-nesting}
$\!\!\! .\,\,$
There is a constant $c>0$ such that,
for any $\eps>0$, there 
exists an integer $\nu\leq 2^{\eta^c } \log (1/ \eps)$ and a 
finite union $K$ of intervals of total length less than $\eps$ such that $Z_\nu= Z_{\nu-1}$,
for any $\delta\in \Omega\setminus K$.
\end{lemma}

Note that $Z_\nu= Z_{\nu-1}$ implies that $Z=Z_\nu$.
Indeed, suppose that $Z_{t+1}\supset Z_t$ for $t\geq \nu$;
then, $f^{t+1}(\mathbf{y})\in D$ but $f^{t}(\mathbf{y})\not\in D$
for some $\mathbf{y}\in \mathbb{S}$; in other words,
$f^{\nu}(\mathbf{x})\in D$ but $f^{\nu-1}(\mathbf{x})\not\in D$
for $\mathbf{x}= f^{t-\nu+1}(\mathbf{y})$, which contradicts
the equality $Z_\nu= Z_{\nu-1}$.

\begin{corollary}\label{MIS-LC}
$\!\!\! .\,\,$    
For $\delta$ almost everywhere in $\Omega$,\footnote{Meaning
outside a subset of $\Omega$ of Lebesgue measure zero.}
every orbit is asymptotically periodic.
\end{corollary}

\proof
The equality $Z= Z_\nu$ implies the eventual periodicity of
the symbolic dynamics. The period cannot exceed the number
of connected components in the complement of $Z$.
Once an  itinerary becomes periodic at time $t_o$ with period $\sigma$,
the map $f^t$ can be expressed locally by matrix powers. Indeed, divide $t-t_o$ by $\sigma$
and let $q$ be the quotient and $r$ the remainder; then, locally,
$f^t= g^{q}\circ f^{t_o+r}$, where $g$ is specified by
a stochastic matrix with a positive diagonal, which implies convergence to a periodic point
at an exponential rate. Finally, apply Lemma~\ref{MIS-nesting} repeatedly, with
$\eps= 2^{-l}$ for $l=1,2,\ldots$ and denote by
$K_l$ be the corresponding union of ``forbidden" intervals. Define
$K^l= \bigcup_{j\geq l}K_j$ and $K^\infty= \bigcap_{l>0} K^l$; then
Leb$(K^l)\leq  2^{1-l}$ and hence Leb$(K^\infty)=0$.
The lemma follows from the fact that any $\delta\in \Omega$ outside of $K^\infty$
lies outside of $K^l$ for some $l>0$.
\hfill $\Box$
\proofend

The corollary states that the set of ``nonperiodic" values of $\delta$ has measure zero
in parameter space. Our result is actually stronger than that. We prove that 
the nonperiodic set can be covered by a Cantor set of Hausdorff dimension strictly less than~1.
The remainder of this section is devoted to a proof of Lemma~\ref{MIS-nesting}.

\subsection{Shift spaces and growth rates}\label{ShiftSpace}

The {\em growth exponent} of a language is defined as  
$\lim_{n\rightarrow \infty}\frac{1}{n}\max_{k\leq n}\log N(k)$,
where $N(k)$ is the number of words of length $k$;
for example, the growth exponent of $\{0,1\}^*$ is 1
(all logarithms taken to the base~2).
The language consisting of all the itineraries of a Markov influence system
forms a {\em shift space} and its growth exponent is the {\em topological entropy}
of its symbolic dynamics~\cite{sternberg10}---not to be confused
with the topological entropy of the {\em MIS} itself.
It can be strictly positive, which is a sign of chaos.
We show that, for a typical system, it is zero, the key fact
underlying periodicity. 
Let $M_1,\ldots, M_T$ be $n$-by-$n$ matrices from
a finite set $\mathcal{M}$ of primitive stochastic rational 
matrices with positive diagonals,
and assume that $\tau(M)< 1/2$ for $M\in \mathcal{M}$;
hence $\tau(M_1\cdots M_k) < 2^{-k}$.
Because each product $M_1\cdots M_k$ is a primitive matrix, it can be expressed as
$\mathbf{1}\bm{\pi}_k^\top + Q_k$ (by Perron-Frobenius), where  
$\bm{\pi}_k$ is its (unique) stationary distribution.\footnote{Positive diagonals
play a key role here because primitiveness is not closed under multiplication:
for example,
$\bigl( \begin{smallmatrix}1 & 1\\ 1 & 0\end{smallmatrix}\bigr)$
and
$\bigl( \begin{smallmatrix}0 & 1\\ 1 & 1\end{smallmatrix}\bigr)$
are both primitive but their product is not.}
If $\bm{\pi}$ is a stationary distribution for a
stochastic matrix $S$, then its $j$-th row $\bm{s}_j$ satisfies
$\bm{s}_j-\bm{\pi}^\top = 
\bm{s}_j-\bm{\pi}^\top  S= \sum_i \bm{\pi}_i( \bm{s}_j -  \bm{s}_i)$;
hence, by the triangular inequality,
$\|\bm{s}_j-\bm{\pi}^\top\|_1\leq  \sum_i \pi_i\| \bm{s}_j -  \bm{s}_i\|_1
\leq 2\tau(S)$. This implies that

\begin{equation}\label{Qdecay}
\begin{cases}
\hspace{.2cm}
M_1\cdots M_k= \mathbf{1}\bm{\pi}_k^\top + Q_k \\
\hspace{.2cm}
\|Q_k\|_\infty\leq 2\tau(M_1\cdots M_k) < 2^{1-k}.
\end{cases}
\end{equation}

\paragraph{Property $\mathbf{U}$.}

Fix a vector $\mathbf{a}\in \mathbb{Q}^n$,
and denote by $M^{\,(\theta)}$ the $n$-by-$m$ matrix
with the $m$ column vectors $M_1\cdots M_{k_i}\, \mathbf{a}$, where
$\theta= (k_1,\ldots, k_m)$ is an increasing sequence of integers in $[T]$.
We say that property $\mathbf{U}$ holds if there 
exists a rational vector $\mathbf{u} = \mathbf{u}(\theta)$ such that $\mathbf{1}^\top\mathbf{u}=1$
and $\mathbf{x}^\top M^{\,(\theta)} \mathbf{u}$ does not depend on the variable
$\mathbf{x}\in \mathbb{S}$.\footnote{Because $\mathbf{x}$ is
a probability distribution, property $\mathbf{U}$ does {\em not} imply that
$M^{\,(\theta)} \mathbf{u}=\mathbf{0}$; for example, we have
$\mathbf{x}^\top \bigl(\mathbf{1}\mathbf{1}^\top \bigr)\, \mathbf{u}
=~1$ for $\mathbf{u}=  \frac{1}{n}\mathbf{1}$.}
Property $\mathbf{U}$ is a quantifier elimination device
for expressing a notion of ``general position" for an {\em MIS}. To see why, consider
a simple statement such as ``the three points $\bigl(x,x^2\bigr)$, 
$\bigl(x+1, (x+1)^2\bigr)$, and $\bigl(x+2, (x+2)^2\bigr)$ cannot be collinear for any
value of $x$." This can be expressed by
saying that a certain determinant polynomial in $x$ is constant.
Likewise, the vector $\mathbf{u}$ manufactures a quantity,
$\mathbf{x}^\top M^{\,(\theta)} \mathbf{u}$, that 
``eliminates" the variable $\mathbf{x}$. Some condition on $\mathbf{u}$ is
needed since otherwise we could pick $\mathbf{u}= \mathbf{0}$.
Note that property $\mathbf{U}$ would be obvious if all the matrices $Q_k$ 
in~(\ref{Qdecay}) were null: indeed, we would have
$\mathbf{x}^\top M^{\,(\theta)} 
= \mathbf{x}^\top \mathbf{1}
      (b_1, \ldots, b_m)
=   (b_1, \ldots, b_m)$,
where $b_i= \bm{\pi}_{k_i}^\top \, \mathbf{a}$.
This suggests that property $\mathbf{U}$ rests on the decaying properties of~$Q_k$.

To see the relevance of general position to the dynamics of an {\em MIS}, consider the iterates
of a small ball through the map $f$. To avoid chaos, it is intuitively obvious
that these iterated images should not fall across discontinuities too often.
Fix such a discontinuity: if we think of the ball as being so small it looks like a point,
then the case we are trying to avoid consists of many points (the ball's iterates) lying on (or near)
a given hyperplane. This is similar to the definition of general position,
which requires that a set of point should not lie on the same hyperplane.

\begin{lemma}\label{MIS-gen-basic}
$\!\!\! .\,\,$
There exists a constant $b>0$ (linear in $n$) such that, given any integer $T>0$
and any increasing sequence $\theta$ in $[T]$ of length at least 
$T^{1-\alpha}/\alpha$, 
property $\mathbf{U}$ holds, where $\alpha \! := \mu^{-b}$ and $\mu$
is the maximum number of bits needed to encode any entry of
$M_k$ for any $k\in [T]$.
\end{lemma}
 
\proof
By choosing $b$ large enough, we can ensure that $T$ 
is as big as we want.
The proof is a mixture of algebraic and combinatorial arguments.
We begin with a Ramsey-like statement about stochastic matrices.

\begin{fact}\label{MIS1}
$\!\!\! .\,\,$
There is a constant $d>0$ such that, if the sequence $\theta$
contains $j_0,\ldots, j_{n}$ with $j_i \geq d\mu j_{i-1}$ for each $i\in [n]$,  
then property $\mathbf{U}$ holds.
\end{fact}

\proof
By~(\ref{Qdecay}),
$\|Q_k\mathbf{a}\|_\infty <c_0 2^{-k}$ for constant $c_0>0$.
Note that $Q_k$ has rational entries over $O(\mu k)$ bits:
the bound follows from the fact that the stationary distribution
$\bm{\pi}_k$ has rational coordinates over $O(\mu k)$ bits;
as noted earlier, the constant factors may depend on $n$.
We write $M^{\,(\theta)}= \mathbf{1}\mathbf{a}^\top \Pi^{\,(\theta)} + Q^{\,(\theta)}$, 
where $\Pi^{\,(\theta)}$ and $Q^{\,(\theta)}$ are the $n$-by-$m$ matrices formed
by the $m$ column vectors $\bm{\pi}_{k_i}$ and $Q_{k_i} \mathbf{a}$, respectively,
for $i\in [m]$; recall that $\theta= (k_1,\ldots, k_m)$.
The key fact is that the dependency on $\mathbf{x}\in \mathbb{S}$ is confined to 
the term $Q^{\,(\theta)}$: indeed, 
\begin{equation}\label{dependX}
\mathbf{x}^\top M^{\,(\theta)}\mathbf{u}
= \mathbf{a}^\top \Pi^{\,(\theta)}\mathbf{u} + \mathbf{x}^\top Q^{\,(\theta)}\mathbf{u}.
\end{equation}
This shows that, in order to satisfy property $\mathbf{U}$,
it is enough to ensure that $Q^{\,(\theta)}\mathbf{u}=0$
has a solution such that $\mathbf{1}^\top \mathbf{u} =1$.
Let $\sigma = (j_0,\ldots, j_{n-1})$.
If $Q^{\, (\sigma)}$ is nonsingular then, because each one of its entries
is a rational over $O(\mu j_{n-1})$ bits, we have
$|\det Q^{\, (\sigma)}|\geq c_1^{\mu j_{n-1}}$, for constant $c_1>0$. 
Let $R$ be the $(n+1)$-by-$(n+1)$ matrix derived from $Q^{\, (\sigma)}$ by adding the column
$Q_{j_n} \mathbf{a}$ to its right and then adding a row of ones at the bottom.
If $R$ is nonsingular, then $R\, \mathbf{u}=(0,\ldots, 0,1)^\top$ has a (unique)
solution in $\mathbf{u}$ and property $\mathbf{U}$ holds (after padding 
$\mathbf{u}$ with zeroes).
Otherwise, we expand the determinant of $R$ along the last column.
Suppose that $\det Q^{\, (\sigma)}\neq 0$.
By Hadamard's inequality, all the cofactors are at most a constant $c_2>0$
in absolute value; hence, for $d$ large enough,
$$
0= |\det R \, |\geq |\det Q^{\, (\sigma)}\, | - nc_2 \|Q_{j_n}\mathbf{a}\, \|_\infty 
\geq c_1^{\mu j_{n-1}} - nc_2 c_0 2^{-j_n}>0.$$
This contradiction implies that $Q^{\, (\sigma)}$ is singular,
so (at least) one of its rows can be expressed as a linear combination of
the others. We form the $n$-by-$n$ matrix $R'$ by
removing that row from $R$, together with the last column,
and setting $u_{j_n}=0$ to rewrite 
$Q^{\,(\theta)}\mathbf{u}=0$ as
$R'  \mathbf{u'}=(0,\ldots, 0,1)^\top$, where
$\mathbf{u'}$ is the restriction of $\mathbf{u'}$ to the columns indexed by $R'$.
Having reduced the dimension of the system by one variable, we can 
proceed inductively in the same way; either we terminate with the
discovery of a solution or the induction runs its course until $n=1$ 
and the corresponding 1-by-1 matrix is null, so that the solution~1 works.
Note that $\mathbf{u}$ has rational coordinates over $O(\mu T)$ bits.
\hfill $\Box$
\proofend

Let $N(T)$ be the largest sequence $\theta$ in $[T]$ such that
property $\mathbf{U}$ does not hold.
Divide $[T]$ into bins $[(d\mu)^k, (d\mu)^{k+1}-1]$ for $k\geq 0$.
By Fact~\ref{MIS1}, the sequence $\theta$ can intersect at most $2n$ of them;
thus, if $T> t_0$, for some large enough $t_0= (d\mu)^{O(n)}$,
there is at least one empty interval in $T$ of length 
$T/(d\mu)^{2n+3}$. This gives us the recurrence $N(T)\leq T$ for $T\leq t_0$
and $N(T)\leq N(T_1)+N(T_2)$, where $T_1+T_2\leq \beta T$, for a positive
constant $\beta= 1- (d\mu)^{-2n-3}$.
The recursion to the right of the empty interval, say, $N(T_2)$, warrants a brief discussion.
The issue is that the proof of Fact~\ref{MIS1} relies crucially on the property that
$Q_k$ has rational entries over $O(\mu k)$ bits---this is needed 
to lower-bound $|\! \det Q^{\, (\sigma)}|$ when it is not 0.
But this is not true any more, because, after the recursion, the columns of
the matrix $M^{\,(\theta)}$ are of the form
$M_1\cdots M_{k} \, \mathbf{a}$, for $T_1+L< k\leq T$, where $L$ is the length
of the empty interval and $T=T_1+L+T_2$. 
Left as such, the matrices use too many bits for the recursion to go through.
To overcome this obstacle,
we observe that the recursively transformed $M^{\,(\theta)}$ can be factored as $AB$, where
$A= M_1\cdots M_{T_1+L}$ and $B$ consists of the
column vectors $M_{T_1+L+1}\cdots M_{k} \, \mathbf{a}$.
The key observation now is that, if
$\mathbf{x}^\top B \, \mathbf{u}$ does not depend on $\mathbf{x}$, then
neither does $\mathbf{x}^\top M^{\,(\theta)}\, \mathbf{u}$,
since it can be written as $\mathbf{y}^\top B \, \mathbf{u}$
where $\mathbf{y}= A^\top \mathbf{x}\in \mathbb{S}$.
In this way, we can enforce property $\mathbf{U}$
while having restored the proper encoding length for the entries of
$M^{\,(\theta)}$.

Plugging in the ansatz $N(T)\leq t_0 T^\gamma$,
for some unknown positive $\gamma<1$, we find by Jensen's inequality that, for all $T>0$,
$N(T)\leq t_0( T_1^\gamma + T_2^\gamma)
\leq t_0 2^{1-\gamma}\beta^\gamma T^\gamma$.
For the ansatz to hold true, we need to ensure that 
$2^{1-\gamma}\beta^\gamma \leq 1$.
Setting $\gamma= 1/(1- \log\beta)<1$ completes the proof of Lemma~\ref{MIS-gen-basic}.
\hfill $\Box$
\proofend

Define $\phi^k (\mathbf{x})  = \mathbf{x}^\top M_{1}\cdots M_{k}$
for $\mathbf{x}\in \mathbb{R}^n$ and $k\leq T$; 
and let $h_\delta \! : \mathbf{a}^\top \mathbf{x}=1+\delta$ be some hyperplane in $\mathbb{R}^n$.
We consider a set of canonical intervals of length $\rho$ (or less):
$\mathcal{D}_\rho = 
      \bigl\{\, [k\rho, (k+1)\rho]\cap \Omega\,|\, k\in \mathbb{Z}\,\bigr\}$,
where $\rho>0$ (specified below),
$\Omega= \omega\mathbb{I}$, $\mathbb{I}\! := [-1,1]$, and $0<\omega<1/2$.
Roughly, the ``general position" lemma below says that, for most $\delta$, the $\phi^k$-images of
any $\rho$-wide cube centered in the simplex $\mathbb{S}$ 
cannot come very near the hyperplane $h_\delta$ for most values of $k\leq T$. 
This may be counterintuitive.
After all, if the stochastic matrices $M_i$ are the identity,
the images stay put, so if the initial cube collides then all of the
images will! The point is that $M_i$ is primitive so it cannot be the identity.
The low coefficients of ergodicity will also play a key role.
The crux of the lemma is that the exclusion set $U$ does not
depend on the choice of $\mathbf{x}\in \mathbb{S}$.
A point of notation: $\alpha$ refers to its use in Lemma~\ref{MIS-gen-basic}.

\begin{lemma}\label{MISexPk}
$\!\!\! .\,\,$
For any real $\rho>0$ and any integer $T>0$, there exists $U\subseteq \mathcal{D}_\rho$
of size $c_T= 2^{O(\mu T)}$, where $c_T$ is independent of $\rho$,
such that, for any $\Delta\in  \mathcal{D}_\rho \!\setminus\! U$
and $\mathbf{x}\in \mathbb{S}$, there are at most $T^{1-\alpha}/\alpha$ 
integers $k\leq T$ such that
$\phi^k (X) \cap h_\Delta\neq \emptyset$, where 
$X=  \mathbf{x}+ \rho\mathbb{I}^n$ and $h_\Delta \! := \bigcup_{\delta\in \Delta} h_\delta$.
\end{lemma}

\proof
In what follows, $b_0,b_1,\ldots$ refer to suitably large positive constants
(which, we shall recall, may depend on $n, \mathbf{a}$, etc).
We assume the existence of more than $T^{1-\alpha}/\alpha$ integers
$k\leq T$ such that $\phi^k(X) \cap h_\Delta\neq \emptyset$ 
for some $\Delta \in \mathcal{D}_\rho$ and 
draw the consequences: in particular, we  infer 
certain linear constraints on $\delta$;
by negating them, we define the forbidden set $U$ and ensure the conclusion of the lemma.
Let $k_1<\cdots<k_m$ be the integers in question, where $m> T^{1-\alpha}/\alpha$.
For each $i\in [m]$, there exists $\mathbf{x}(i)\in X$ and $\delta_i\in \Delta$ such that  
$|\, \mathbf{x}(i)^\top M_{1}\cdots M_{k_i}\mathbf{a} - 1 - \delta_i\, |  \leq \rho$.
Note that $|\delta_i - \delta|\leq \rho$ for some $\delta\in \Delta$ common to all $i\in [m]$.
By the stochasticity of the matrices, it follows that 
$ |\, \bigl(\mathbf{x}(i)- \mathbf{x}\bigr)^\top M_{1}\cdots M_{k_i}\mathbf{a}\, | \leq
b_0\,  \rho$; hence
$|\, \mathbf{x}^\top M_{1}\cdots M_{k_i}\mathbf{a} - 1 - \delta\, |  \leq (b_0+2) \rho$.
By  Lemma~\ref{MIS-gen-basic}, there is a rational vector $\mathbf{u}$ 
such that $\mathbf{1}^\top\mathbf{u}=1$ 
and $\mathbf{x}^\top M^{\,(\theta)} \mathbf{u}= \psi(M^{\,(\theta)}, \mathbf{a})$ 
does not depend on the variable $\mathbf{x}\in \mathbb{S}$;
on the other hand, 
$|\, \mathbf{x}^\top M^{\,(\theta)} \mathbf{u} - (1+\delta) \, |\leq b_1\rho$.
Two quick remarks: 
(i) the term $1+\delta$ is derived from $(1+\delta)\mathbf{1}^\top\mathbf{u}=1+\delta$;
(ii) $b_1\leq (b_0+2) \| \mathbf{u}\|_1$, where $\mathbf{u}$ is a rational 
over $O(\mu T)$ bits.
We invalidate the condition on $k_1,\ldots, k_m$ by keeping $\delta$
outside the interval 
$\psi(M^{\,(\theta)}, \mathbf{a}) -1 +b_1\rho \mathbb{I}$, which rules out
at most $2(b_1+1)=  2^{O(\mu T)}$ intervals from $\mathcal{D}_\rho$.  Repeating this for all
sequences $(k_1,\ldots, k_m)$ raises the number of forbidden intervals
by a factor of at most~$2^T$.  
\hfill $\Box$
\proofend

\paragraph{Topological entropy.}

We identify the family $\mathcal{M}$ with the set of all matrices of the form
$S_{\! 1}\cdots S_{\! k}$, for $\eta\leq k\leq 3\eta$ ($\eta= \eta_\Omega$),
where the matrix sequence $S_{\! 1}, \ldots, S_{\! k}$ matches 
some element of $L_\Omega^k$.
By definition of the ergodic renormalizer, any $M\in \mathcal{M}$ is primitive
and $\tau(M)< 1/2$; furthermore,  
both $\mu$ and $\log|\mathcal{M}\,|$ are in $O(\eta)$.
Our next result shows that the topological entropy of the shift space of itineraries vanishes.

\begin{lemma}\label{MIS-GR}
$\!\!\! .\,\,$
For any real $\rho>0$ and any integer $T>0$,
there exist $t_\rho= O(\eta |\log\rho \,|)$
and an exclusion set $V\subseteq \mathcal{D}_\rho$ of size 
$d_T= 2^{O(T)}$ such that,
for any $\Delta\in \mathcal{D}_\rho \!\setminus\! V$,
any integer $t\geq t_\rho$, and any $\sigma\in L_\Delta^t$,
$\log \, \bigl| \, \bigl\{ \, \sigma' \,|\, 
  \sigma\cdot \sigma'\in L_\Delta^{t+ T}\,\bigr\}\,\bigr|\leq
 \eta^{b}  T^{1-\eta^{-b}}$,
for constant $b>0$.
\end{lemma}
 
\proof
In the lemma, $t_\rho$ (resp. $d_T$) is independent of $T$
(resp. $\rho$). The main point is that the exponent of $T$ 
is bounded away from~1. 
We define $V$ as the union of the sets $U$ formed by applying 
Lemma~\ref{MISexPk} to each one of the hyperplanes $h_\delta$ 
involved in $\mathcal{P}$ and every possible sequence of
$T$ matrices in $\mathcal{M}$. 
This increases $c_T$ to $2^{O(\eta T)}$.
Fix $\Delta\in  \mathcal{D}_\rho \!\setminus\! V$
and consider the (lifted) phase space
$\mathbb{S}\times \Delta$ for the dynamical system induced by the map
$f_{\uparrow} \!: 
\bigl(\mathbf{x}^\top,\delta)\mapsto (\,  \mathbf{x}^\top S\!(\mathbf{x}), \delta \,\bigr)$.
The system is piecewise-linear with respect to the polyhedral
partition $\mathcal{P}_{\uparrow}$ of $\mathbb{R}^{n+1}$  
formed by treating $\delta$ as a variable in $h_\delta$.
Let $\Upsilon_t$ be a continuity piece for $f_{\uparrow}^t$, ie,
a maximal region of $\mathbb{S}\times \Delta$ over which
the $t$-th iterate of $f_{\uparrow}$ is linear.
Reprising the argument leading to~(\ref{Qdecay}), 
any matrix sequence $S_{\!1},\ldots, S_{\! t}$ matching
an element of $L_\Delta^t$ is such that
$S_{\!1}\cdots S_{\! t} = \mathbf{1}\bm{\pi}^\top + Q$,
where
\begin{equation}\label{SprodBound}
\|Q\|_\infty < 2^{2-t/\eta}.
\end{equation}
Thus there exists
$t_\rho= O(\eta |\log\rho \,|)$ 
such that, for any $t\geq t_\rho$, $f_{\uparrow}^t(\Upsilon_t)\subseteq 
(\mathbf{x}+\rho\mathbb{I}^n)\times \Delta$,
for some $\mathbf{x}= \mathbf{x}(t,\Upsilon_t)\in \mathbb{S}$.
Consider a nested sequence 
$\Upsilon_1\supseteq \Upsilon_2\supseteq \cdots$.
Note that 
$\Upsilon_1$ is a cell of $\mathcal{P}_{\uparrow}$, 
$f_{\uparrow}^{k}(\Upsilon_{k+1})\subseteq f_{\uparrow}^k(\Upsilon_k)$,
and $S_{\! l}$ is the stochastic matrix
used to map $f_{\uparrow}^{l-1}(\Upsilon_l)$ 
to $f_{\uparrow}^{l}(\Upsilon_l)$ (ignoring the $\delta$-axis).
We say there is a {\em split} at $k$ if  $\Upsilon_{k+1} \subset \Upsilon_k$,
and we show that, given any $t\geq t_\rho$, there are only
$O(\eta  T^{1-\alpha}/\alpha)$ splits between $t$
and $t+ \eta T$, where $\alpha= \eta^{-b}$, for constant $b$.\footnote{We 
may have to scale $b$ up by a constant factor since
$\mu= O(\eta)$ and, by Lemma~\ref{MIS-gen-basic}, $\alpha= \mu^{-b}$.}
We may confine our attention to splits caused by the same
hyperplane $h_\delta$ since $\mathcal{P}$ features only
a constant number of them.
Arguing by contradiction, we assume the presence of
at least $6 \eta  T^{1-\alpha}/\alpha$ splits,
which implies that at least $N \! := 2 T^{1-\alpha} /\alpha$ of those splits 
occur for values of $k$ at least $2\eta$ apart.
This is best seen by
binning $[t+1,t+\eta T]$ into $T$ intervals of length $\eta$
and observing that at least $3N$ intervals must feature splits. In fact,
this proves the existence of $N$ splits
at positions separated by a least two consecutive bins.
Next, we use the same binning to produce the matrices $M_1,\ldots, M_T$, where
$M_j= S_{\! t+1+(j-1)\eta}\cdots S_{\! t+ j\eta}$.

Suppose that all of the $N$ splits occur for values 
$k$ of the form $ t+ j \eta$. In this case, a straightforward
application of Lemma~\ref{MISexPk} is possible:
we set $X\times \Delta= f_{\uparrow}^t(\Upsilon_t)$
and note that the functions $\phi^k$ are all products of matrices from the family
$\mathcal{M}$, which happen to be $\eta$-long products.
The number of splits, $2 T^{1-\alpha} /\alpha$, exceeds the number allowed 
by the lemma and we have a contradiction.
If the splits do not fall neatly at endpoints of the bins,
we use the fact that $\mathcal{M}$ includes matrix products of any
length between $\eta$ and $3\eta$.
This allows us to reconfigure the bins so as to form a sequence
$M_1,\ldots, M_T$ with the splits occurring at the endpoints:
for each split, merge its own bin with the one to its left
and the one to its right (neither of which contains a split) and use the split's position
to subdivide the resulting interval into two new bins; we leave
all the other bins alone.\footnote{We note the possibility
of an inconsequential decrease in $T$ caused by the merges.
Also, we can now see clearly why Lemma~\ref{MISexPk} 
is stated in terms of the slab $h_\Delta$ and not the 
hyperplane $h_\delta$.
This allows us to express splitting caused 
by the hyperplane $\mathbf{a}^\top \mathbf{x} =1+\delta$
in lifted space $\mathbb{R}^{n+1}$.}
This leads to the same contradiction, which implies 
the existence of fewer than $O( \eta  T^{1-\alpha}/\alpha )$ splits
at $k\in [t,  t+ \eta T]$; hence the same
bound on the number of strict inclusions in 
the nested sequence $\Upsilon_t\supseteq \cdots \supseteq \Upsilon_{t+ \eta T}$.
The set of all such sequences forms a tree
of depth $\eta T$, where each node has 
at most a constant number of children 
and any path from the root has $O(\eta  T^{1-\alpha}/\alpha)$
nodes with more than one child. Rescaling $T$ to $\eta T$
and raising $b$ completes the proof.
\hfill $\Box$
\proofend

\subsection{Proof of Lemma~\ref{MIS-nesting}}

We show that the nonperiodic $\delta$-intervals can be covered by
a Cantor set of Hausdorff dimension less than one.
All the parameters below refer to Lemma~\ref{MIS-GR}.
We fix $\eps>0$ and $\Delta\in \mathcal{D}_\rho \!\setminus\! V$
and assume that $\delta\in \Delta$.
%
%
% NOTE !!!!!    Why T has to be so big?  Because we must have $\eta^b< T^{\eta^{-b}$.
%                        Also, when checking the math,  leave log(1/\rho) as such as long as possible
%
Set $T= 2^{\eta^{2b}}$, $\rho= \eps/(2d_T)$, and
$\nu= t_\rho + kT$, where $k= c \eta \log(1/\rho)$ for
a large enough constant $c>0$.
Since $t_\rho= O(\eta |\log\rho \,|)$
and $d_T= 2^{O(T)}$, we have
\begin{equation}\label{nuBound}
\nu= 2^{\eta^{O(1)}} \log(1/\eps) .
\end{equation}
Let $M$ be the matrix $S_{\!1}\cdots S_{\! \nu}$,
where $S_{\!1},\ldots, S_{\! \nu}$ the matrix sequence
matching an element of $L_\Delta^\nu$.
By~(\ref{SprodBound}),
$\text{diam}_{\ell_\infty} (\mathbb{S}\, M ) \leq 2^{2-\nu/\eta}$.
There exists a point $\mathbf{x}_M$
such that, given any point $\mathbf{y}\in \mathbb{S}$ 
whose $\nu$-th iterate $f^\nu(\mathbf{y})=  \mathbf{z}^T$ is specified by
the matrix $M$, that is, $\mathbf{z}^T= \mathbf{y}^T M$, we have 
$\|\mathbf{x}_M - \mathbf{y}\|_\infty\leq  2^{2-\nu/\eta}$.
Consider a discontinuity $h_\delta: \mathbf{a}_i^\top \mathbf{x} =1+\delta$
of the system. Testing which side of it the point $\mathbf{z}$
lies is equivalent to checking the point $\mathbf{x}_M$ instead 
with respect to $h_{\delta'}$ for 
some $\delta'$ that differs from $\delta$ by $O(2^{-\nu/\eta})$.
It follows that adding an interval of length $O(2^{-\nu/\eta})$
to the exclusion set $V$ ensures that all the $\nu$-th iterates 
$f^\nu(\mathbf{y})$ (specified by $M$)
lie strictly on the same side of $h_\delta$ for all $\delta\in \Delta$.
Repeating this for every string $L_\Delta^\nu$ 
and every $\Delta\in \mathcal{D}_\rho \!\setminus\! V$ increases the length
covered by $V$ from its original $d_T\rho = \eps/2$ to at most
$d_T\rho + O( |L_\Delta^\nu| 2^{-\nu/\eta}/\rho) < \eps$.
This last bound follows from the consequence of Lemma~\ref{MIS-GR} that
$\log |L_\Delta^\nu| \leq k \eta^{b}  T^{1-\eta^{-b}} + O(t_\rho)$.
Thus, for any $\delta\in \Omega$ outside a set of intervals
covering a length less than $\eps$, no $f^\nu(\mathbf{x})$ lies
on a discontinuity. It follows that, for any such $\delta$, we have $Z_{\nu}= Z_{\nu-1}$ 
and, by~(\ref{nuBound}), the proof is complete.
\hfill $\Box$
\proofend

\section{Applications}\label{apps}

We can use the renormalization and bifurcation analysis techniques
developed above to resolve several important families of {\em MIS}.  
We discuss two simple cases here.

\subsection{Irreducible systems}

A Markov influence system is called {\em irreducible} 
if the Markov chain $g(\mathbf{x})$ 
is irreducible for all $\mathbf{x}\in \mathbb{S}$; given the assumed
presence of self-loops, each chain is ergodic.
All the digraphs $g(\mathbf{x})$ of an irreducible {\em MIS} are strongly connected.
Every step from the beginning sees growth in
the cumulant until it is a clique. 
To see why, assume by contradiction that the cumulant fails to grow 
at an earlier step, ie, 
$\prod_{< k}\mathbf{g}  = \prod_{\leq k}\mathbf{g}$ for $k\leq m_1$, where
$m_1$ is the smallest index for which $\prod_{\leq m_1}\mathbf{g}$ is a clique.
If so, then $g_k$ is a subgraph of $\text{\em{tf}}\,(\prod_{< k}\mathbf{g} )$.
Because the latter is transitive and it has in $g_k$ a strongly connected subgraph
that spans all the vertices, it must be a clique;
therefore $\prod_{< m_1}\mathbf{g}$ is a clique, 
which contradicts our assumption. Because $m_1\leq n$,
each product of at most $n$ graphs is a clique, so
its coefficient of ergodicity is at most $1-\gamma$, for $\gamma>0$.
Since the number of distinct matrices is finite, their positive entries 
are uniformly bounded away from zero; hence
$\eta_\Omega$ is bounded from below by a positive
constant (which may depend on $n$).
By Lemma~\ref{MIS-nesting}
and Corollary~\ref{MIS-LC}, we conclude with:

\begin{theorem}\label{MIS-irreducible}
$\!\!\! .\,\,$    
Typically, every orbit of an irreducible
Markov influence system is asymptotically periodic.
\end{theorem}

\subsection{Weakly irreducible systems}

We strengthen the previous result by assuming
a fixed partition of the vertices
such that each digraph $g(\mathbf{x})$ consists of disjoint
strongly connected graphs defined over the subsets $V_1,\ldots, V_l$ of the 
partition. This means that no edges ever join two distinct $V_i, V_j$
and, within each $V_i$, the graphs are always strongly connected with self-loops.
Irreducible systems correspond to the case $l=1$.
What makes weak irreducibility interesting is that the systems are
not simply the union of independent irreducible systems. 
Indeed, communication flows among states
in two ways: (i) directly, vertices collect information
from incoming neighbors to update their states; and (ii) indirectly, via the
polyhedral partition $\mathcal{P}$,
the sequence of graphs for $V_i$
may be determined by the current states within other groups $V_j$. 
In the extreme case, we can have the co-evolution of 
two systems $V_1$ and $V_2$, each one depending
entirely on the other one yet with no links between the two of them.
If the two subsystems were independent,
their joint dynamics could be expressed as a direct sum
and resolved separately. This cannot be done, in general,
and the bifurcation analysis requires some modifications.

By using topological renormalization, we can partition the vertex set $[n]$
into $l$ subsets and deal with each one separately.
For the bifurcation analysis, Fact~\ref{MIS1} relies on the rank-$l$ expansion
\begin{equation*}
\mathbf{x}^\top M^{\,(\theta)}
=  \sum_{i=1}^l \,\, 
\Bigl(\, \sum_{j\in V_i}x_j \,\Bigr)\, \mathbf{a}^\top \Pi_i^{\,(\theta)} 
          + \mathbf{x}^\top Q^{\,(\theta)}\, ,
\end{equation*}
where 
(i) $M_1\cdots M_k= \sum_{i=1}^l \mathbf{1}_{|V_i} \bm{\pi}_{i,k}^\top + Q_k$;
(ii) all vectors $\mathbf{1}_{|V_i}$ and $\bm{\pi}_{i,k}$ have support in $V_i$;
(iii)  $Q_k$ is block-diagonal and $\|Q_k\|_\infty <2^{1-k}$;
(iv) $\Pi_i^{\,(\theta)}$ and $Q^{\,(\theta)}$ are formed, respectively,
by the column vectors $\bm{\pi}_{i, k_j}$ and $Q_{k_j} \mathbf{a}$ for $j\in [m]$,
with $\theta= (k_1,\ldots, k_m)$.
Property $\mathbf{U}$ no longer holds, however:
if $l>1$, indeed, it is no longer true that 
$\mathbf{x}^\top M^{\,(\theta)} \mathbf{u}$ is independent of the variable
$\mathbf{x}\in \mathbb{S}$. The dependency is confined to the sums
$s_i:= \sum_{j\in V_i}x_j$ for $i\in [l]$.
The key observation is that these sums are time-invariant.
We assume them to be rational and
we redefine the phase space
as the invariant manifold $\prod_{i=1}^l \bigl( s_i \, \mathbb{S}^{|V_i|-1}\bigr)$.
The rest of the proof mimics the irreducible case,
whose conclusion therefore still applies.

\begin{theorem}\label{MIS-weakly-irreducible}
$\!\!\! .\,\,$    
Typically, every orbit of a weakly irreducible
Markov influence system is asymptotically periodic.
\end{theorem}

\section{Hyper-Torpid Mixing and Chaos}\label{chaos}

Among the {\em MIS} that converge to a single stationary distribution,
we show that the mixing time can be super-exponential.
Very slow clocks can be designed in the same manner:
the {\em MIS} is periodic with a period of length 
equal to a tower-of-twos.
The creation of new timescales is what most distinguishes
{\em MIS} from standard Markov chains.   As we mentioned earlier,
the system can be chaotic. We prove all of these claims below.

\subsection{A super-exponential mixer}

How can reaching a fixed point distribution take so long?  Before we answer this
question formally, we provide a bit of intuition. Imagine having 
three unit-volume water reservoirs $A, B, C$ alongside a clock that
rings at noon every day.
Initially, the clock is at 1pm and $A$ is full while $B$ and $C$ are empty.
Reservoir $A$ transfers half of its content to $B$ and repeats this each hour
until the clock rings noon. At this point, reservoir $A$ empties into $C$ the little water
that it has left and $B$ empties its content into $A$.
At 1pm, we resume what we did the day before at the same hour, ie,
$A$ transfers half of its water content to $B$, etc.  This goes on until some day, 
the noon hour rings and 
reservoir $C$ finds its more than half full. (Note that the water level of $C$ rises 
by about $10^{-3}$ every day.) 
At this point, both $B$ and $C$ transfer all their water back to $A$, so that
at 1pm on that same day, we are back to square one. The original 12-step clock has been
extended into a new clock of period roughly 1,000. 
The proof below shows how to simulate this iterative process with an {\em MIS}.

\begin{theorem}\label{TorpidMixing}
$\!\!\! .\,\,$
There exist Markov influence systems 
that mix to a stationary distribution in time 
equal to a tower-of-twos of height linear in the number of states.
\end{theorem}

\proof
We construct an {\em MIS} with a periodic orbit
of length equal to a power-of-twos of height proportional to $n$;
this is the function $f(n)= 2^{f(n-1)}$, with $f(1)=1$.
It is easy to turn it into one with
an orbit that is attracted to a stationary
distribution (a fixed point) with an equivalent mixing rate,
and we omit this part of the discussion.
Assume, by induction, that we have a Markov influence system $M$ cycling through 
states $1,\ldots, p$, for $p\geq 4$. We build another
one with period at least $2^p$ by adding a ``gadget"
to it consisting of a graph over the vertices $1,2,3$ 
with probability distribution $(x,y,z)\in \mathbb{S}$.
We initialize the system by placing $M$ in state 1 (ie, 1pm in our clock example)
and setting $x=1$.
The dynamic graph is specified by these rules:
\begin{enumerate}
\item
Suppose that $M$ is in state $1,\ldots, p-1$.
The graph has the edge $(1,2)$, which is assigned 
probability $1/2$, as is the self-loop at~1. There are self-loops at 2 and 3.
\item
Suppose that $M$ is in state $p$.
If $z\leq 1/2$, then the graph has the edge $(2,1)$ and $(1,3)$, both
of them assigned probability $1$, with one self-loop at 3.
If $z> 1/2$, then the graph has the edge $(2,1)$ and $(3,1)$, both
of them assigned probability $1$, with one self-loop at 3.
\end{enumerate}

Suppose that $M$ is in state 1 and that $y=0$ and $z\leq 1/2$.
When $M$ reaches state $p-1$, then $x= (1-z)2^{1-p}$
and $y= (1-z)(1-2^{1-p})$. Since $z\leq 1/2$,
the system cycles back to state 1 with the updates
$z\leftarrow x+z$ and $x\leftarrow y$.
Note that $z$ increases by a number between
$2^{-p}$ and $2^{1-p}$.
Since $z$ begins at 0, such increases will occur consecutively at least
$p2^p/4\geq 2^p$ times, before $x$ is reset to $1$.
The construction on top of $M$ adds three new vertices
so we can push this recursion roughly $n/3$ times to produce a Markov influence system
that is periodic with a period of length equal to a tower-of-twos of height roughly~$n/3$.

We need to tie up a few loose ends.
The construction needs to recognize state $p-1$
by a polyhedral cell; in fact, any state will do.
The easiest choice is state 1, which corresponds to $x\geq 1$ (to express it
as an inequality). 
The base case of our inductive construction consists of
a two-vertex system of period $p=4$ with initial distribution $(1,0)$.
If $x>  2^{1-p}$, the graph has an edge from 1 to 2 and a self-loop at 1,
both of them assigned probability $1/2$; else
an edge from 2 to 1 given probability 1 to reset the system.
Finally, the construction assumes probabilities summing up to 1 within each 
of the $\lfloor (n-2)/3\rfloor+1$ gadgets, which is clearly wrong:
we fix this by dividing the probability weights equally among the gadget and adjusting
the linear discontinuities appropriately. 
\hfill $\Box$
\proofend

\subsection{Chaos}

We give a simple 5-state construction with chaotic symbolic dynamics.
The idea is to build an {\em MIS} that simulates the 
classic baker's map.

\begin{equation*}
{\small 
A= \frac{1}{3}
\begin{pmatrix}
\, 2 & 1 & 0 & 0 & 0 \\
\, 0 & 1 & 2 & 0 & 0 \\
\, 0 & 0 & 3 & 0 & 0 \\
\, 0 & 0 & 0 & 2 & 1 \\
\, 0 & 0 & 0 & 0 & 3 \,
\end{pmatrix} }
\hspace{.3cm}
\text{if} 
\hspace{.2cm}
x_1+ x_2  > x_4
\hspace{.7cm}
\text{and}
\hspace{.6cm}
\small{
B= \frac{1}{3}
\begin{pmatrix}
\, 1 & 0 & 2 & 0 & 0 \\
\, 1 & 2 & 0 & 0 & 0 \\
\, 0 & 0 & 3 & 0 & 0 \\
\, 0 & 0 & 0 & 2 & 1 \\
\, 0 & 0 & 0 & 0 & 3 \,
\end{pmatrix} }\hspace{.2cm}
\text{else},
\end{equation*}
for $\mathbf{x} \in \mathbb{S}^4$.
We focus our attention on
$\Sigma=\bigl\{ \, (x_1,x_2, x_4)\,|\, 0<x_1\leq x_4/2\leq x_2<x_4\, \bigr\}$,
and easily check that it is an invariant manifold.
At time $0$, we fix $x_4=1/4$ and $x_5=0$; at all times, of course,
$x_3=1-x_1-x_2-x_4-x_5$.
The variable $y\! : = (2x_2-x_4)/(2x_1-x_4)$ 
is always nonpositive over $\Sigma$.
It evolves as follows:
\begin{equation*}
y\leftarrow 
\begin{cases}
\, \frac{1}{2}(y+1)  \hspace{1.345cm} \text{ if  $y< -1$} \\
\,  \frac{2y}{y+1}    \hspace{2cm} \, \text{ if $-1\leq y\leq 0$.}
\end{cases}
\end{equation*}
Writing $z= (y+1)/(y-1)$, we note that $-1 \leq z < 1$ and it evolves
according to 
$z\mapsto 2z+1$ if $z\leq 0$, and $z\mapsto 2z-1$ otherwise,
a map that conjugates with the baker's map
and is known to be chaotic~\cite{devaney}.

\bigskip

\subsection*{Acknowledgments}

I wish to thank Maria Chudnovsky and Ramon van Handel for helpful comments.

%%% \newpage

%%% \appendix
%%% \section*{Appendix}

\end{document}